# Atomic models of non-stoichiometric layered diborides $M_{1-x}B_2$ (M = Mg, Al, Zr and Nb) from first principles

I.R. Shein, A.L. Ivanovskii [*]

*Institute of Solid State Chemistry, Ural Branch of the Russian Academy of Sciences, 620041, Ekaterinburg, Russia*

**Abstract**

Two atomic models of non-stoichiometric metal diborides $M_{1-x}B_2$ are now assumed: (i) the presence of cation vacancies and (ii) the presence of "super-stoichiometric" boron which is placed in cation vacancy site. We have performed first principle total energy calculations using the VASP-PAW method with the generalized gradient approximation (GGA) for the exchange-correlation potential in a perspective to reveal the trends of their possible stable atomic configurations depending on the type of M cations (M = Mg, Al, Zr or Nb) and the type of the defects (metal vacancies *versus* metal vacancies occupied by "super-stoichiometric" boron in forms of single atoms, dimers $B_2$ or trimers $B_3$). Besides we have estimated the stability of these non-stoichiometric states (on the example of magnesium-boron system) as depending on the possible synthetic routes, namely *via* solid state reaction method, as well as in reactions between *solid* boron and Mg *vapor*; and between these reagents in *gaseous phase*. We demonstrate that the non-stoichiometric states such as $B_2$ and $B_3$ placed in metal sites may be stabilized, while the occupation of vacancy sites by single boron atoms is the most unfavorable.

---

[*] Corresponding author.
   *E-mail address:* shein@ihim.uran.ru (I.R. Shein).

# 1. Introduction

The majority of metal borides at equilibrium conditions have very narrow ranges of homogeneity [1-5] and the influence of non-stoichiometry on their properties has been ignored for a long time. The discovery [6] of superconductivity ($T_C \sim 39K$) in $MgB_2$ gave rise to the intensive investigations of metal diborides ($MB_2$) and related materials, reviews [7-15]. In particular, it was revealed that besides other infringements of ideal crystal structure of $MB_2$ phases such as substitutional impurities, dislocations, mesoscopic inhomogeneity, grain boundaries, various radiation defects, etc., the deviation of compositions of diborides from ideal stiochiometry affects $T_C$, see [16-19].

By the present time, the presence of cation non-stoichiometry (*i.e.* B/M > 2) for the diborides of *s, p, d* metals is a well established fact. Except for $Mg_{1-x}B_2$ (x ~ 0.04-0.05) [16-19], the significant deviation from the ideal stiochiometry was observed for aluminum diboride ($Al_{1-x}B_2$, x ~ 0.1-0.11) [20,21] and especially for diborides of group V *d* metals ($Nb_{1-x}B_2$ and $Ta_{1-x}B_2$, x = 0.01-0.5) [22].

The non-stoichiometry changes the structural, physical and chemical properties of these borides, but the effect may be opposite for the different borides. For example, it is well known (reviews [10,12]) that the superconductivity in $MgB_2$ is caused by the strong electron-phonon interaction with the participation of the $E_g$ phonon modes and the electronic σ - bands arising from the planar nets of boron atoms, so any infringements of crystal structure including the cation non-stoichiometry have the negative influence on $T_C$ (for $Mg_{1-x}B_2$ (x ~ 0.05) the value of $T_C$ is reduced at ~ 2-3K [16-19]). On the contrary, for ideal stoichiometric niobium diboride $NbB_{2.0}$ the superconductivity is absent down to $T_C < 2.5K$ [23], whereas $T_C$ at about 9K was reported for the cation-deficient samples $Nb_{1-x}B_2$ [22, 24-27]. The atomic picture of the boron-rich compositions $M_{1-x}B_2$ remains debatable and the excess of boron in $M_{1-x}B_2$ is attributed to the presence of metal vacancies [28-31], or the M-vacancies are supposed to be occupied by "super-stoichiometric" boron in form of $B_2$ dimers [32]. The experimental arguments for each of these models are discussed, for example, in Ref. [32].

In this paper, we report the results of *ab-initio* study of the non-stoichiometric *s,p,d* metal diborides $M_{1-x}B_2$ with a goal to predict their possible stable atomic structure depending on the M cation (M = Mg, Al, Zr or Nb) and the type of defects (metal vacancies and metal vacancies occupied by single boron atoms, dimers $B_2$ or trimers $B_3$).
In addition, as it is well known [1-15], that the synthesis conditions may have a significant influence on the microstructure, superconducting and other properties of metal diborides, we have estimated the stability of these non-stoichiometric phases depending on the possible synthetic routes, namely *via* solid state reaction method, as well as in reactions between *solid* boron and Mg *vapor* and between these reagents in *gaseous phase*.

## 2. Models and method

The considered metal diborides $MB_2$ adopt the layered $AlB_2$-like crystal structure (space group P6/*mmm*). It is a simple hexagonal lattice of close-packed metal layers alternating with graphite-like B layers in a sequence ..*AHAHAH*.. perpendicularly to the *c* direction, as depicted in Fig. 1. Here the boron atoms are arranged at the corners of a hexagon with three nearest neighbor B atoms in each plane. The metal atoms are located at the center of the B hexagons between adjacent boron layers. Thus, the M and B atoms have the $[MB_{12}M_8]$ and $[BM_6B_3]$ coordination polyhedra, respectively. Each metal center has $D_{6h}$ symmetry, i.e. 12 boron atoms at the vertices of a hexagonal prism. In addition, the central metal atom is coordinated also by 8 metals through the faces of the $B_{12}$ prisms. We simulated the nonstoichiometric diborides by the 24 atoms supercells $\{M_8B_{16}\}$, i.e. (2×2×2) unit cells. The cation vacancy ($V^M$) was modeled by the removal of M atom and such supercell

{$M_7V^MB_{16}$} corresponds to the formal stoichiometry $M_{0.875}B_2$. The boron super-stoichiometric diborides were simulated using the supercells {$M_7V^M(B_{16}+B)$}, {$M_7V^M(B_{16}+B_2)$} and {$M_7V^M(B_{16}+B_3)$}, where substitutional single boron atom, dimer $B_2$ or trimer $B_3$ were placed in cation vacancy site, see Fig.1. These supercells describe the nominal compositions $M_{0.875}B_{2.125}$, $M_{0.875}B_{2.250}$ and $M_{0.875}B_{2.375}$, respectively.

Our calculations were performed within the projector augmented wave (PAW) method [33] in formalism of density functional theory as implemented in the VASP package [34-36]. We used the VASP PAW potentials with three valence electrons for boron ($2s^22p^1$), two for Mg ($3s^23p^0$), three for Al ($3s^23p^1$) and $2+n$ ($4d^n5s^2$); $n = 2$ and 3 for Zr and Nb, respectively. The exchange-correlation functional with the generalized gradient approximation (GGA) in the PBE form [37] was employed. We used a well-converged energy cutoff of 400 eV for plane-wave expansion of the PAW's, and the Blöchl's tetrahedron method [38] for the Brillouin zone integrations. The relaxation effects due to the presence of cation vacancy and "super-stoichiometric" boron were taken into account, and the relaxation of all internal coordinates was performed until the forces were less than 0.005 eV/Å.

## 3. Results and discussion

To examine the effect of non-stoichiometry on the stability of $MB_2$ phases, their formation energies ($E_{form}$) were estimated according to the formal reactions (i) for complete phases: $2B + M \rightarrow MB_2$; (ii) for non-stoichiometric $M_{1-x}B_2$ phases with cation vacancies: $2B + (1-x)M \rightarrow M_{1-x}B_2$; and (iii) for non-stoichiometric $M_{1-x}B_{2+y}$ phases with "super-stoichiometric" boron: $(2+y)B + (1-x)M \rightarrow M_{1-x}B_{2+y}$) as:

$$E_{form}^{MB_2} = E_{tot}(MB_2) - \{E_{tot}(M_{cond}) + 2\,E_{tot}(B_{cond});$$

$$E_{form}^{M_{1-x}B_2} = E_{tot}(M_{1-x}B_2) - \{(1-x)\,E_{tot}(M_{cond}) + 2\,E_{tot}(B_{cond})\};$$

$$E_{form}^{M_{1-x}B_{2+y}} = E_{tot}(M_{1-x}B_{2+y}) - \{(1-x)\,E_{tot}(M_{cond}) + (2+y)\,E_{tot}(B_{cond})\};$$

where $E_{tot}(M_xB_y)$, $E_{tot}(M_{cond})$ and $E_{tot}(B_{cond})$ are the calculated total energies $E_{tot}$ for borides as well as for pure metals (*hcp* Mg, *fcc* Al, *hcp* Zr and *bcc* Nb) and the most stable allotrope of crystalline boron ($\alpha$-$B_{12}$). Thus, here $E_{form}$ may be treated as the total energy difference between boride and a mechanical mixture of their constituent components. This means that if the formation energy $E_{form}$ is negative (positive), the boride phase is stable (unstable) relative to a separation into the elemental metal and elemental boron.

For all stoichiometric borides we obtained $E_{form} < 0$ (Table 1). In agreement with the experimental data and the previous theoretical estimations [1-5,10,12,30,31,39] our results show that the negative $E_{form}$ for complete $MB_2$ phase decreases as $E_{form}(ZrB_2) > E_{form}(NbB_2) > E_{form}(AlB_2) > E_{form}(MgB_2)$. This means that $ZrB_2$ is the most stable phase, which possesses among the considered metal diborides the maximal thermal and strength characteristics.

These results can be understood in terms of the chemical bonding and the band filling in $MB_2$, see also [10,12,39,40]. The valence band of $MB_2$ is determined mainly by the B $2p$ states, which form four $\sigma(2p_{x,y})$ and two $\pi(2p_z)$ bands, where the $p_\sigma$ orbitals are responsible for the strongest covalent B-B bonds inside planar boron sheets. For $ZrB_2$, the optimum condition of the band occupation is satisfied when all the bonding states are occupied and all the antibonding states are vacant. An increase or decrease in the electron concentration for others diborides leads to the occupation of antibonding bands or depletion of the bonding bands that results in a decrease of stability of these phases. Note also that our estimations of $E_{form}$ (for example, for $ZrB_2$ (-291.2 kJ/mol) and $NbB_2$ (-188.9 kJ/mol)) agree reasonably with the experimental thermodynamic data, the enthalpies for $ZrB_2$ and $NbB_2$ are equal -318 kJ/mol [41] and -180 − -197 kJ/mol [42,43], respectively.

The results for non-stoichiometric $M_{1-x}B_2$ and $M_{1-x}B_{2+y}$ phases are listed in Table 1. For all non-stoichiometric phases their formation energies are less, than for ideal $MB_2$ systems, *i.e.* the occurrence of metal vacancies break some M-M and M-B bonds and decrease the stability of hexagonal borides. The minimum stability is found for the $M_{0.875}B_{2.125}$ phases, where non-stoichiometry was simulated by the single boron atom in vacancy position, *i.e.* the occupation of vacancy sites by single boron atoms is most unfavorable.

There is a strong dependence of the most stable non-stiohiometric states on the metal, see Table 1. For $Mg_{1-x}B_2$, the most favorable state corresponds to the $B_3$ trimers at the Mg vacancies, whereas for $Al_{1-x}B_2$, $Zr_{1-x}B_2$ and $Nb_{1-x}B_2$ the formation of metal vacancy is more preferably. It should be noted that (i) for $Al_{1-x}B_2$ all states containing the metal vacancies occupied by "super-stoichiometric" boron (in forms of single atoms, dimers $B_2$ or trimers $B_3$) have the positive formation energies and (ii) for $Zr_{1-x}B_2$ and $Nb_{1-x}B_2$ their formation energies for states with metal vacancies and metal vacancies with the $B_2$ dimers are comparable. Probably, these differences for *s*, *p* or *d* metal diborides are related to the features of rearrangement of inter-atomic bonds in the vicinity of these defects.

Above we analyzed the stability of structural states of non-stoichiometric diborides relative to pure metals and crystalline boron α-$B_{12}$ and in this way we simulated the formation of these compounds from the *condensed* reagents, *i.e.* their synthesis *via* solid state reaction method. Meanwhile some diborides, for example $MgB_2$, may be prepared *via* alternative synthetic routes involving a reaction between *solid* boron and Mg *vapor*. Other possible route is a reaction between the reagents in *gaseous phase*, which is realized in various films preparation *in situ* techniques. Thus, we estimated the formation energies ($E'_{form}$) for $MgB_2$ and above non-stoichiometric $Mg_{1-x}B_2$ and $Mg_{1-x}B_{2+y}$ phases *via* reaction between *solid* boron and Mg *vapor* as:

$$E'_{form}{}^{MgB_2} = E_{tot}(MgB_2) - \{E_{tot}(Mg_{at}) + 2\, E_{tot}(B_{cond});$$
$$E'_{form}{}^{Mg_{1-x}B_2} = E_{tot}(Mg_{1-x}B_2) - \{(1-x)\, E_{tot}(Mg_{at}) + 2\, E_{tot}(B_{cond});$$
$$E'_{form}{}^{Mg_{1-x}B_{2+y}} = E_{tot}(Mg_{1-x}B_{2+y}) - \{(1-x)\, E_{tot}(Mg_{at}) + (2+y)\, E_{tot}(B_{cond})\};$$

where $E_{tot}(Mg_{at})$ and $E_{tot}(B_{cond})$ are the total energies for free Mg atom and crystalline boron α-$B_{12}$. In turn, the formation energies ($E''_{form}$) for non-stoichiometric phases *via* reaction between *gaseous* boron and Mg were determined as:

$$E''_{form}{}^{MgB_2} = E_{tot}(MgB_2) - \{E_{tot}(Mg_{at}) + 2\,E_{tot}(B_{at})\};$$
$$E''_{form}{}^{Mg_{1-x}B_2} = E_{tot}(Mg_{1-x}B_2) - \{(1-x)\,E_{tot}(Mg_{at}) + 2\,E_{tot}(B_{at})\};$$
$$E''_{form}{}^{Mg_{1-x}B_{2+y}} = E_{tot}(Mg_{1-x}B_{2+y}) - \{(1-x)\,E_{tot}(Mg_{at}) + (2+y)\,E_{tot}(B_{at})\};$$

where $E_{tot}(Mg_{at})$ and $E_{tot}(B_{at})$ are the total energies for the free Mg and boron atoms. The analysis of the change of formation energies for non-stoichiometric diborides relative to the corresponding values for complete $MgB_2$ ($\Delta E_{form} = E_{form}(MgB_2) - E_{form}(Mg_xB_y)$) presented in Table 2 leads to the following main conclusions:

(i) $\Delta E_{form}$ are positive, i.e. the formation of all defect types is energetically unfavorable in comparison with the complete phase and the non-equilibrium process are necessary to synthesize the non-stoichiometric diborides.

(ii) Various types of defects may be preferable depending on the aggregative states of reagents (crystalline or gaseous): for solid-state route the lowest value of $\Delta E_{form}$ corresponds to the composition with the $B_3$ trimers inside M-vacancy ($Mg_{0.875}B_{2.250}$), whereas in synthetic routes with gaseous reagents the most preferable are the systems with "pure" cationic vacancies ($Mg_{0.875}B_2$).

(iii) For any routes the presence in vacancy sites of the single boron atom ($Mg_{0.875}B_{2.125}$) or molecule $B_2$ ($Mg_{0.875}B_{2.375}$) is unfavorable.

Let's discuss the possible influence of non-stoichiometric states on the electronic properties of magnesium diboride. The total and site-projected *l*-decomposed densities of states (DOS) for complete $MgB_2$ as well as for $Mg_{0.875}B_2$, $Mg_{0.875}B_{2.125}$, $Mg_{0.875}B_{2.250}$ and $Mg_{0.875}B_{2.375}$ calculated for the equilibrium geometries of these phases are shown in Figs. 2 and 3. For $MgB_2$ our calculations well reproduce the earlier results, see [9,10,12].

The total densities of states for $Mg_{0.875}B_2$ as well as for $Mg_{0.875}B_{2.125}$ are similar to DOS for stoichiometric phase, though the narrow peaks DOSs peak near $E_F$ are formed, Fig. 2. The DOS at the Fermi level $N(E_F)$ for these non-stoichiometric phases varies less ±2 % in comparison with $MgB_2$.

More changes there are in DOS for $Mg_{0.875}B_{2.250}$ and $Mg_{0.875}B_{2.375}$ with $B_2$ and $B_3$ "molecules" placed in the Mg-vacancy sites. The additional low-energy DOS peaks (A, Fig 2) are formed mainly by the B 2$s$ states of $B_2$ or $B_3$ "molecules", Fig. 3. Note also that for these systems with strong B-B covalent bonds for $B_2$ or $B_3$ "molecules" their DOSs shapes resembles boron DOS for stoichiometric $MgB_2$, where also strong B-B interactions in boron lattice take place, see also [9-12]. In result for $MgB_2$, as well as for $Mg_{0.875}B_{2.250}$ and $Mg_{0.875}B_{2.375}$ the dominant role of boron 2p states at the Fermi level ($N^{B\ 2p}(E_F) > N^{B\ 2s}(E_F)$) occur, while for $Mg_{0.875}B_{2.125}$ with single boron atom in metal vacancy the opposite situation is obtained: $N^{B\ 2p}(E_F) < N^{B\ 2s}(E_F)$, see Fig. 3.

The value of $N(E_F)$ sharply decreases from 0.70 states/eV ($MgB_2$) to 0.61 and 0.40 states/eV as going from $MgB_2$ to $Mg_{0.875}B_{2.250}$ and $Mg_{0.875}B_{2.375}$, respectively. Within the framework of the BCS model ($T_C \sim <\omega> \exp\{f(\lambda)\}$, where $<\omega>$ represents the averaged phonon frequency, the coupling constant $\lambda = N(E_F) <I^2>/<M\omega^2>$, where $<I^2>$ is an averaged electron-ion matrix element squared, M is an atomic mass), one of the reasons why $T_C$ of non-stoichiometric B-rich magnesium diboride decreases [16-19] may be connected with the lowering of $N(E_F)$ for the systems with $B_3$ trimers at Mg-site.

## 4. Conclusions

In summary, the first principle total energy calculations using the VASP-PAW method with the generalized gradient approximation (GGA) for the exchange-correlation potential have been performed to reveal the possible stable atomic states of non-stoichiometric (B/M >2) metal diborides $M_{1-x}B_2$, namely the presence of cation vacancies *versus* the presence of "super-stoichiometric" boron (in forms of single atoms, dimers $B_2$ or trimers $B_3$) placed in cation vacancy sites.

Our main findings are:

(i) The formation of all types of non-stoichiometric states are energetically unfavorable - in comparison with the complete $MB_2$, in this sense, the importance of a non-equilibrium process to synthesize the non-stoichiometric compositions of metal diborides should be concluded.

(ii) The occupation of vacancy sites by single boron atoms for all diborides is most unfavorable.

(iii) The trends of formation of stable atomic configurations for non-stoichiometric $M_{1-x}B_2$ are depending strongly from the type of M sublattice (M = Mg, Al, Zr or Nb). For $Mg_{1-x}B_2$ the most favorable state include $B_3$ trimers inside Mg vacancies, whereas for all others diborides: $Al_{1-x}B_2$, $Zr_{1-x}B_2$ and $Nb_{1-x}B_2$ the formation of metal vacancy is more preferably.

(iv) For $Al_{1-x}B_2$ all states containing metal vacancies occupied by "super-stoichiometric" boron (in forms of single atoms, dimers $B_2$ or trimers $B_3$) adopts the positive formation energies, *i.e.* are unstable; and for $Zr_{1-x}B_2$ and $Nb_{1-x}B_2$ their formation energies for states with metal vacancies and metal vacancies with $B_2$ dimers are comparable.

(v) Our estimations of the stability of non-stoichiometric states for $Mg_{1-x}B_2$ as depending from the possible synthetic routes, have shown that depending of the aggregative states of reagents (crystalline or gaseous) the various types of defects may be preferable: for solid-state route the lowest value of $\Delta E_{form}$ correspond the composition with $B_3$ trimers placed inside Mg-vacancy, whereas in synthetic routes with gaseous reagents the most preferable are the systems with "pure" cationic vacancies. Besides, the electronic structures for various states of non-stoichiometric magnesium diboride are calculated and analyzed.

**Table 1**
Formation energies ($\Delta E_{form}$, eV/f.u.) for $M_{1-x}B_2$, $M_{1-x}B_2$ and $Mg_{1-x}B_{2+y}$ phases (where M = Mg, Al, Zr and Nb) according to VASP-PAW-GGA calculations.

| phase / parameter | $-E_{form}$ * | phase / parameter | $-E_{form}$ |
|---|---|---|---|
| $MgB_2$ | 0.423 | $AlB_2$ | 0.137 |
| $Mg_{0.875}B_2$ | 0.105 | $Al_{0.875}B_2$ | **0.143** |
| $Mg_{0.875}B_{2.125}$ | -0.148 | $Al_{0.875}B_{2.125}$ | -0.322 |
| $Mg_{0.875}B_{2.25}$ | 0.115 | $Al_{0.875}B_{2.25}$ | -0.249 |
| $Mg_{0.875}B_{2.375}$ | **0.154** | $Al_{0.875}B_{2.375}$ | -0.493 |
| phase / parameter | $-E_{form}$ | phase / parameter | $-E_{form}$ |
| $ZrB_2$ | 3.018 | $NbB_2$ | 1.958 |
| $Zr_{0.875}B_2$ | **2.301** | $Nb_{0.875}B_2$ | **1.759** |
| $Zr_{0.875}B_{2.125}$ | 1.915 | $Nb_{0.875}B_{2.125}$ | 1.115 |
| $Zr_{0.875}B_{2.25}$ | 2.213 | $Nb_{0.875}B_{2.25}$ | 1.340 |
| $Zr_{0.875}B_{2.375}$ | 1.996 | $Nb_{0.875}B_{2.375}$ | 1.060 |

* Relatively to the crystalline boron ($\alpha$-$B_{12}$) and the corresponding "pure" metal.

**Table 2**
Formation energies ($\Delta E_{form}$, eV/f.u.) for $Mg_{1-x}B_2$ and $Mg_{1-x}B_{2+y}$ phases in various synthetic routes according to VASP-PAW-GGA calculations.

| phase / parameter * | $\Delta E_{form}$ ** | $\Delta E'_{form}$ | $\Delta E''_{form}$ |
|---|---|---|---|
| $Mg_{0.875}B_2$ | 0.318 | **0.119** | **0.506** |
| $Mg_{0.875}B_{2.125}$ | 0.571 | 0.506 | 2.585 |
| $Mg_{0.875}B_{2.25}$ | 0.308 | 0.496 | 1.485 |
| $Mg_{0.875}B_{2.375}$ | **0.269** | 0.457 | 0.610 |

* Relatively to the corresponding values for complete phase $MgB_2$; i.e. for example: $\Delta E_{form}(Mg_{0.875}B_2) = E_{form}(MgB_2) - E_{form}(Mg_{0.875}B_2)$;

** $\Delta E_{form}$, $\Delta E'_{form}$ и $\Delta E''_{form}$ – formation energies from: condensed Mg and boron, condensed boron and gaseous Mg and from gaseous Mg and boron, respectively, see text.

**Figures:**

Fig. 1. (*Color online*). Fragment of the crystal structures of the layered AlB$_2$-like diborides MB$_2$ and the atomic models of non-stoichiometric (B-rich) states of these phases: 1 – the presence of M - vacancy V$^M$ (supercell ячейка {M$_7$V$^M$B$_{16}$}, formal composition M$_{0.875}$B$_2$) and the presence of "super-stoichiometric" boron (●) as: 2 – single atom, 3 - dimer B$_2$ or 4 - trimer B$_3$ placed in M-vacancy sires/ The corresponding supercells and formal compositions are: {M$_7$V$^M$(B$_{16}$+B)}, {M$_7$V$^M$(B$_{16}$+B$_2$)}, {M$_7$V$^M$(B$_{16}$+B$_3$)} and M$_{0.875}$B$_{2.125}$, M$_{0.875}$B$_{2.250}$ and M$_{0.875}$B$_{2.375}$, respectively.

Fig. 2. Total densities of states for; 1 - complete MgB$_2$ and non-stoichiometric 2- Mg$_{0.875}$B$_2$, 3 - Mg$_{0.875}$B$_{2.125}$, 4 - Mg$_{0.875}$B$_{2.250}$ and 5 - Mg$_{0.875}$B$_{2.375}$.

Fig. 3. Partial boron 2*s* (full lines) and 2*p* (dotted lines) densities of states for: 1- complete MgB$_2$ and boron, B$_2$ and B$_3$ "molecules" in Mg vacancy for non-stoichiometric diborides 2 - Mg$_{0.875}$B$_{2.125}$, 3 - Mg$_{0.875}$B$_{2.250}$ and 4 - Mg$_{0.875}$B$_{2.375}$.

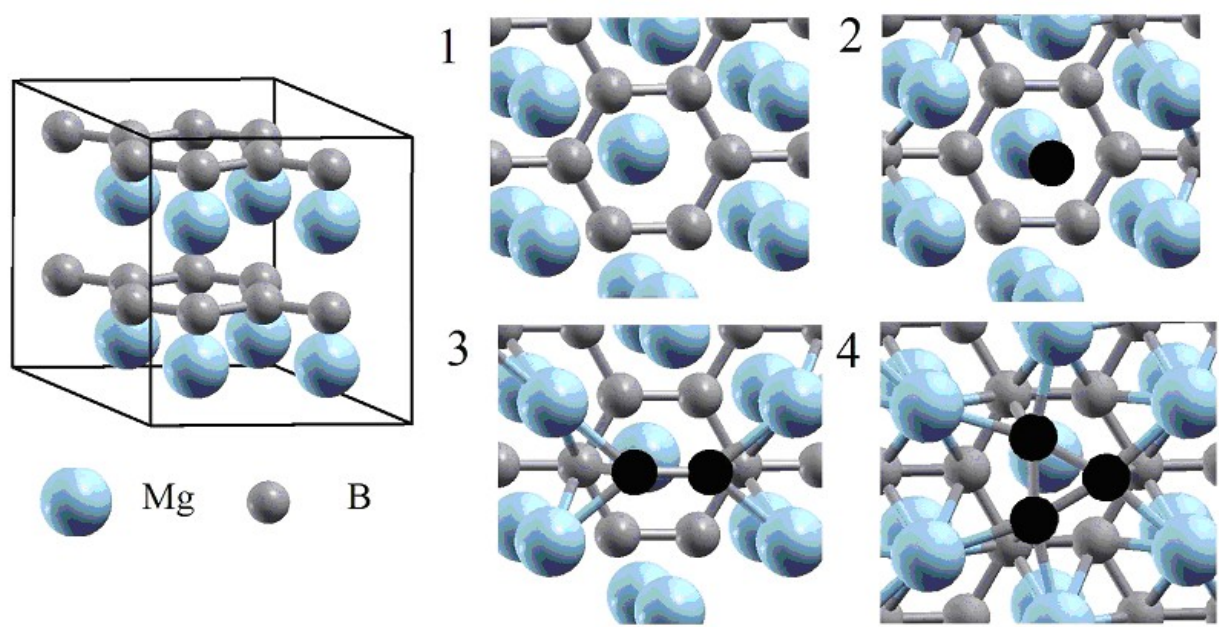

Fig. 1. Shein et al.

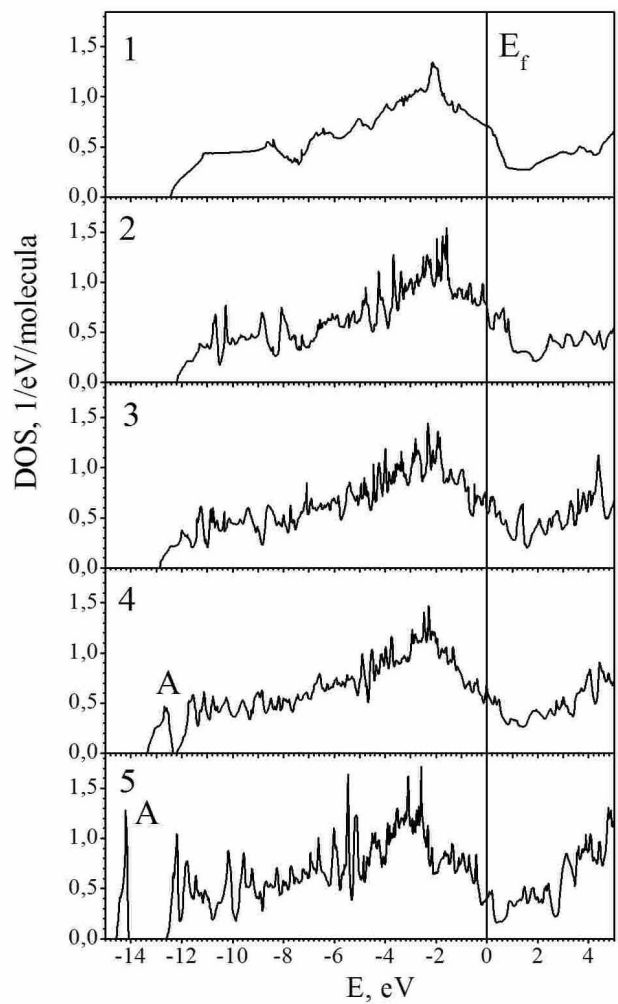

Fig. 2. Shein et al.

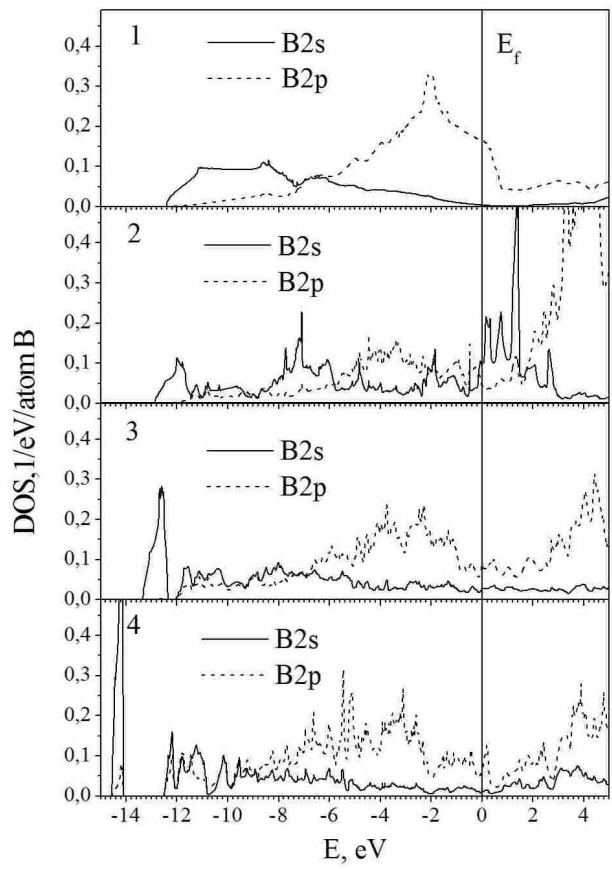

Fig. 3. Shein et al.